\documentclass{PoS}

\font\twelvebb=msbm12
\font\tenbb=msbm10
\font\sevenbb=msbm7
  \newfam\bbfam
  \def\bb{\fam\bbfam\twelvebb}
  \textfont\bbfam=\twelvebb
  \scriptfont\bbfam=\tenbb
  \scriptscriptfont\bbfam=\sevenbb
\font\twelveeusm=eusm10 scaled 1200
\font\teneusm=eusm10
  \newfam\eusmfam
  \def\eusm{\fam\eusmfam\twelveeusm}
  \textfont\eusmfam=\twelveeusm
  \scriptfont\eusmfam=\teneusm
  \scriptscriptfont\eusmfam=\scriptfont\eusmfam
\font\twelvefrak=eufm10 scaled 1200
\font\tenfrak=eufm10
  \newfam\frakfam
  \def\frak{\fam\frakfam\twelvefrak}
  \textfont\frakfam=\twelvefrak
  \scriptfont\frakfam=\tenfrak
  \scriptscriptfont\frakfam=\scriptfont\frakfam
\def\sqr#1#2{{\vcenter{\hrule height.#2pt
   \hbox{\vrule width.#2pt height#1pt \kern#1pt
      \vrule width.#2pt}
   \hrule height.#2pt}}}

\def\bsqr#1#2{{\vrule width #1pt height#2pt}}
\def\bsquare{{\mathchoice\bsqr66\bsqr66\bsqr33\bsqr33}}
\def\badbreak{\penalty1000}

\def\sgn{\mathop{\rm sgn}}                  
\def\Z{{\bb Z}}                             
\newcommand{\cE}{{\cal E}}                  
\newcommand{\fI}{{\frak I}}                 
\newcommand{\cC}{{\cal C}}                  
\newcommand{\qbar}{{\bar q}}                
\newcommand{\Gbar}{{\bar G}}                
\newcommand{\bcC}{{\bar \cC}}               

\title{Dominance of Sign Geometry and the Homogeneity of the Fundamental Topological Structure}

\ShortTitle{Dominance of Sign Geometry and the Homogeneity of the Fundamental Topological\ldots}

\author{\speaker{Ivan Horv\'ath}%
         \thanks{The bulk of data used in this work has been generated at the Center for 
                 Computational Sciences of the University of Kentucky. We thank Mridupawan
                 Deka, Terry Draper, Devdatta Mankame and Jianbo Zhang for help with the data
                 collection at various stages of this project. A computational grant from 
                 SciDAC that was partially used for the required data generation is 
                 acknowledged. We thank U.-J.Wiese for a remark concerning the behavior
                 of considered correlation functions at long distances and to Sumit Das and 
                 Peter Weisz for discussions.}\\
        University of Kentucky, Lexington, KY, USA\\
        E-mail: \email{horvath@pa.uky.edu}}

\author{Andrei Alexandru\\
        The George Washington University, Washington, DC, USA\\
        E-mail: \email{aalexan@gwu.edu}}

\author{Thomas Streuer\\
        University of Regensburg, Regensburg, Germany\\
        E-mail: \email{thomas.streuer@desy.de}}

\abstract{We propose and support the possibility that the shape of topological density 
  2--point function in pure--glue QCD is crucially, and possibly entirely, 
  determined by the space--time folding (geometry) of the double--sheet $\,$ 
  sign--coherent $\,$ structure of Ref. [1], while the distribution of topological density 
  within individual sheets only determines the overall magnitude of the correlator at finite 
  physical distances. A specific manifestation of this, discussed here, 
  is that the shape of the correlation function (encoding e.g. the masses of pseudoscalar 
  glueballs) 
  is reproduced upon the replacement $q(x) \rightarrow \sgn(q(x))$, i.e. by considering 
  the double sheet of the same space--time geometry but with constant 
  magnitude of topological density. Combined with previous results on the fundamental
  topological structure, this suggests that a collective degree of freedom describing topological 
  fluctuations of QCD vacuum can be viewed as a global space-filling 
  {\em homogeneous} double membrane. Selected possibilities for practical uses of this 
  are discussed.}  

\FullConference{The XXVI International Symposium on Lattice Field Theory \\
                 July 14 - 19, 2008\\
                 Williamsburg, Virginia, USA}

\begin{document}

\noindent {\bf 1. The Context.} Discovery of sign--coherent 
topological structure in equilibrium configurations of 
lattice--regularized QCD~\cite{Hor03A} and the subsequent demonstration 
that this order is of dynamical origin~\cite{Ale05A} opened the door 
for systematic inquiry into the nature of configurations dominating 
the QCD path integral~\cite{Hor06fram}. Conceptual innovations associated 
with these developments have their origin in the fact that the space--time 
structure is detected directly in equilibrium ensembles and using local 
composite operators which, among other things, has two important 
implications. 
(1) The information on the space--time order in QCD vacuum so obtained 
is free of a priori assumptions and subjective manipulations  
({\em Bottom--Up approach}~\cite{Hor06fram}) thus putting the associated 
line of research on a solid ground. (2) The space--time structure identified 
in this way incorporates features at all scales in the continuum limit. 
Consequently, the association of these features with underlying 
physics is not limited to usual ``vacuum structure'' low energy 
manifestations (condensates, string tension) but extends to arbitrary 
property described by the theory 
({\em fundamental structure})~\cite{Hor03A,Hor06fram}.

In the general framework of~\cite{Hor06fram} that we 
follow, it is emphasized that the scale dependence of phenomena in QCD
has to be reflected in the way we describe the structure of 
typical configurations (vacuum structure). Indeed, while the fundamental 
structure carries complete information on the theory, it is implicitly 
understood that we are simultaneously considering an infinite tower of 
{\em effective structures}, labeled by momentum scale $\Lambda$. 
In the effective structure the fluctuations up to length scale 
$1/\Lambda$ are averaged out from the fundamental 
structure~\cite{Hor06fram}. An individual effective structure does not 
carry complete information about the theory, but the association of 
its space--time features with physics at the corresponding scale 
$\Lambda$ is expected to capture this physics most efficiently. 
Loosely speaking, effective structures represent the fundamental order 
at varying resolutions (scales of physics). Here we focus on the 
fundamental structure in topological density.

The chief rationale of inquiries into the structure of typical QCD 
configurations is the expectation that such information can eventually be 
turned into an improved qualitative and quantitative control over the theory. 
In case of fundamental topological structure, the underlying order manifests
itself via existence of extended sign--coherent lower--dimensional regions 
(``sheets'') with strong spatial correlation between the regions
of opposite sign~\cite{Hor03A}. More precisely, the following properties of
the sign--coherent structure have been advocated. 
{\em (i) Lower dimensionality}~\cite{Hor03A}, meaning that it is impossible 
(in average sense) to embed a 4-d sign-coherent ball of finite physical 
radius into the sign--coherent topological structure of QCD (see 
also~\cite{Boy06A,Ilg07A}). {\em (ii) Inherent globality}~\cite{Hor03A,Hor05A}, 
meaning that it is not possible to break the coherent structure into localized 
pieces without compromising the physics. In fact, there are 
only two global oppositely charged sheets closely following each other 
(``double--sheet''). {\em (iii) Space--filling nature}~\cite{Hor03A}, meaning 
that even though the structure is locally lower-dimensional, it nevertheless 
fills a macroscopic fraction of 4-d space--time and is thus geometrically 
analogous to space--filling curves. (For additional references related 
to fundamental topological structure, see~\cite{other_fun}.)

In this work, we propose a new geometric property of 
the double--sheet structure that has direct connection to QCD dynamics. 
In particular, we will show that the shape of the 2--point function is insensitive 
to inhomogeneities within the sign--coherent regions present in typical 
configurations of lattice--regularized ensembles. In other words, the presence 
of such inhomogeneities 
is due to an unphysical noise, not affecting correlations at finite physical 
distances. Rather, at the level of fundamental structure, the dynamics appears 
to be encoded (possibly entirely) in the space--time folding of homogeneous 
objects -- in the {\em geometry} of the double--sheet sign--coherent structure.

\begin{figure}
\begin{center}
    \vskip -0.350in
    \centerline{
    \hskip -0.00in
    \includegraphics[width=12.0truecm,angle=0]{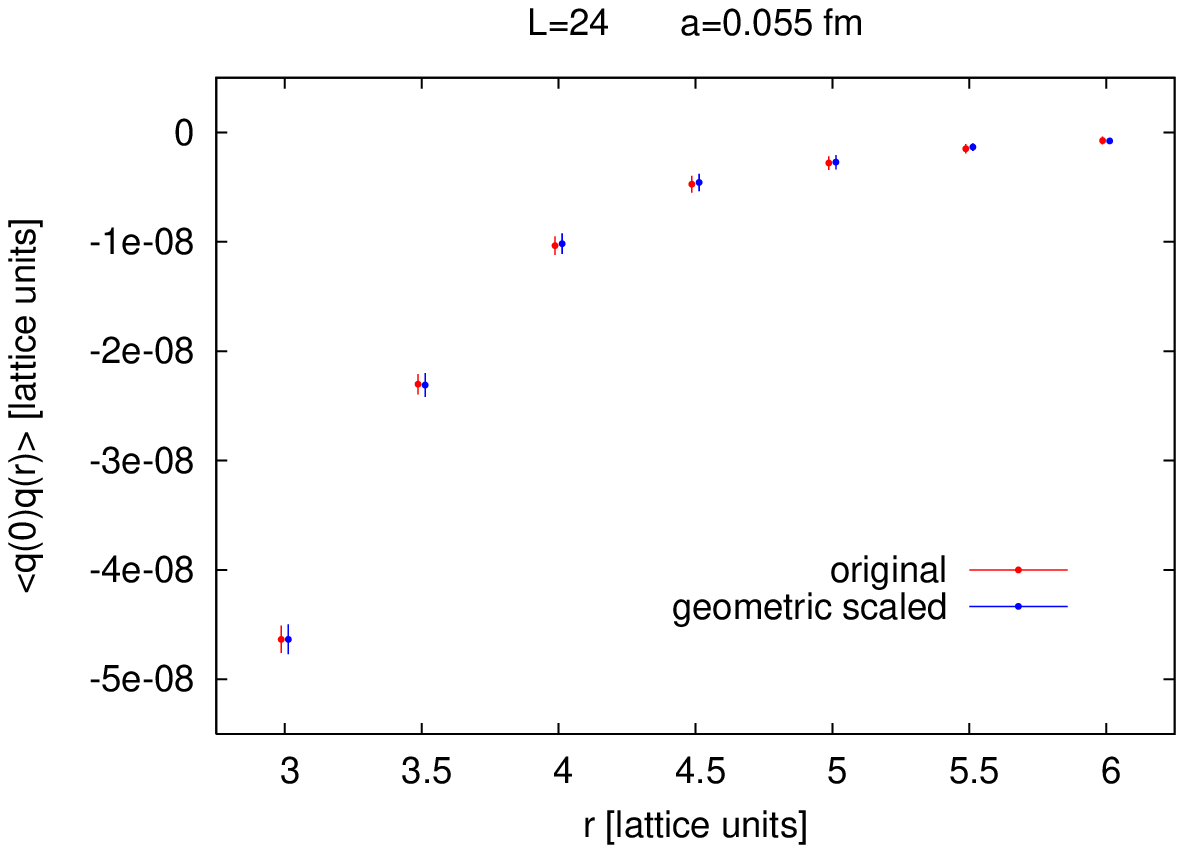}
     }
    \vskip 0.05in
    \centerline{
    \hskip -0.00in
    \includegraphics[width=12.0truecm,angle=0]{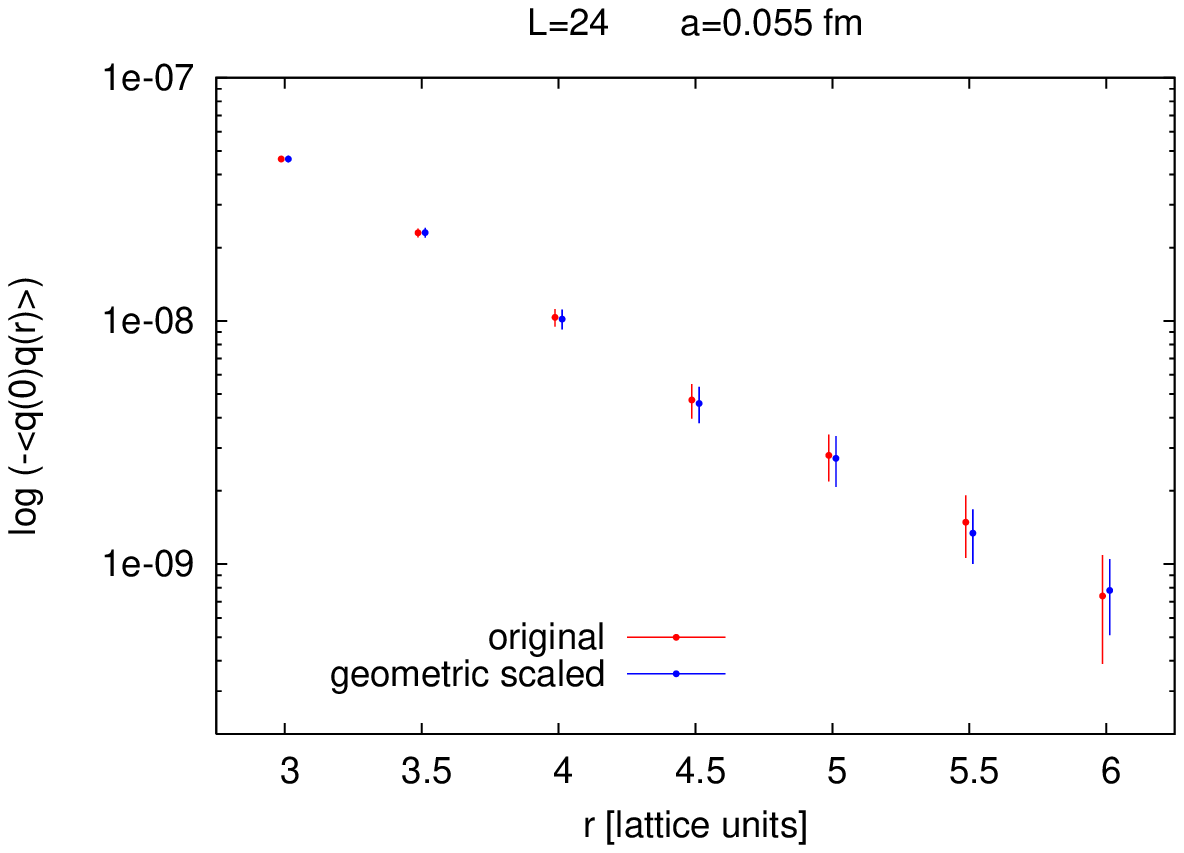}
     }
     \vskip -0.00in
     \caption{The results for the original, $G(r)$, and the geometric,
              $\Gbar(r)$, correlators on $24^4$ lattice in pure--glue
              lattice gauge theory (Iwasaki action) at $a=$ 0.055 fm.
              $\Gbar(r)$ was rescaled to coincide with $G(r)$ at
              $r=3$.}
     \label{fig:bas_point}
     \vskip -0.4in 
\end{center}
\end{figure}

\noindent {\bf 2. The Basic Observation.} Original results 
put forward in this work are based on the outcome of the following
numerical experiment. Consider an ensemble of topological density
configurations corresponding to Wilson's lattice gauge theory   
at some cutoff. With every configuration 
$\cC \equiv \{q(x) , \; x \in \Z^4\}$ associate a new configuration 
$\bcC(\cC)$ defined via
\begin{equation}
   \bcC \,\equiv\, 
   \{ \, sq(x) \equiv \sgn \bigl( q(x) \bigr) , \; q(x) \in \cC \,\}
   \label{eq:10}  
\end{equation}
where $\sgn (y)$ is the sign function. Thus, the configuration $\bcC$ 
has field values $\pm 1$ and possesses the same sign--coherent regions 
as $\cC$. Now, consider the two--point functions of the original
$G(x) \equiv \langle q(x) q(0) \rangle$ and of the associated 
$\Gbar (x) \equiv \langle sq(x) sq(0) \rangle$ 
ensembles. Is there a definite relation between $G(x)$ and $\Gbar(x)$?
Note that we will frequently refer to barred entities as ``geometric'' 
(i.e. geometric correlator, geometric ensemble) since the only information 
kept from the original configuration is the space--time geometry of sign--coherent 
regions. 

In Fig.~\ref{fig:bas_point} we show the result of such calculation
on $24^4$ lattice at lattice spacing $a=$ 0.055 fm, using 
overlap--based definition of topological density. Since the overall 
magnitudes of the two correlators will evidently be very different, 
we have rescaled the geometric one such that at 
$|x| \equiv r = 3$ they become equal, i.e. 
$\Gbar(r) \longrightarrow  \Gbar(r)G(3)/\Gbar(3)$. On the bottom plot,
we show the results on a logarithmic scale to better discern points at 
larger distances. As can be seen quite clearly, the original and 
the rescaled geometric 2--point functions appear identical within 
statistical errors. In fact the agreement is better than statistical,
indicating that, already at this physical volume, sign--coherent geometry 
drives correlations at the level of individual configurations.  

There are couple of points that we would like to note here. 
Firstly, the feature uncovered above only holds in the 
``negative part'' of the lattice correlator, i.e. in the region where 
it can approximate the behavior at finite physical distances, and not 
in the positive core. Indeed, we will not propose that lattice 
correlators $G(x)$ and $\Gbar(x)$ are equal up to rescaling. Rather, 
we will conjecture the possibility that they define the same continuum 
limit at non--zero physical distances. Secondly, we emphasize that
the currently available data~\cite{Hor08B} is supportive of the scenario with 
geometric dominance being satisfied at all, rather than just at asymptotically 
large physical distances.

\noindent {\bf 3. Specifics and Formalization.} The numerical work on which
we draw our conclusions is based on pure--glue SU(3) lattice ensembles defined 
by the Iwasaki action and specified in Table~\ref{ensemb_tab}. The referenced 
lattice spacings are obtained from string tension ($\sqrt{\sigma}=\,$450 MeV), 
and the physical volume is 
kept fixed at $V_p=3$ fm$^4$. For definition of topological density 
operator we use overlap Dirac matrix~\cite{Neu98BA} based on Wilson--Dirac 
kernel with standardly defined parameters $r=1$ and $\rho=26/19$. 
This setup was used in the original work~\cite{Hor03A} as well as in all 
the followups by the Kentucky group. Implementation of matrix--vector operation 
needed to evaluate topological density 
$ q(x) \;=\; \frac{1}{2\rho} \, \mbox{\rm tr} \,\gamma_5 \, D(x,x)$  
is described in Ref.~\cite{Chen03}.             

The correlators shown in Fig.~\ref{fig:bas_point} are coarse--grained over 
the distance of half a lattice spacing. By this we mean that the samples for  
correlations with a given point were accumulated within the spherical shells 
of thickness one half, centered at that point. More precisely, let us define 
a sequence of points and the associated sets (intervals), namely
\begin{equation}
   r_k \,=\, \cases{
      \; 0 \; &  \mbox{\rm for $k=0$} \cr
      \; \frac{k}{2}+\frac{1}{2} \; &  \mbox{\rm for $k=1,2,\ldots$}
   }  \qquad
   {\fI}_k \,=\, \cases{
      \; \{\,0\,\} \; &  \mbox{\rm for $k=0$} \cr
      \; [\,r_k-\frac{1}{4},r_k+\frac{1}{4}\,) 
      \; &  \mbox{\rm for $k=1,2,\ldots$} \cr
       }
    \label{eq:20}
\end{equation}
The coarse--grained correlation function $G(r_k,{\cal C})$ on a given configuration
${\cal C}$ is defined as
\begin{equation}
     G(r_k, {\cal C}) \,=\, \langle \, q(x) q(y) \, 
     \rangle_{\,{\cal C}, \,|x-y| \in {\fI}_k}
     \label{eq:30}
\end{equation}
Continuum limits of the (ensemble--averaged) standard and coarse--grained correlators 
are the same, but the latter behaves more smoothly which is why we use it here. 
All conclusions presented here are independent of this choice. In what follows we will 
denote the coarse--grained correlator simply as $G(r)$ with the above discrete values 
of $r$ implicitly understood.

The results shown in Fig.~\ref{fig:bas_point} correspond to the ensemble $\cE_4$ 
with the ensembles at coarser lattice spacings behaving similarly. To quantify
the trends and to formalize the proposed observation, we need a suitable measure.
Equality of functions up to a multiplicative constant is naturally expressed in terms
of a normalized overlap (scalar product). For continuum real--valued functions 
$f_1$, $f_2$ on the interval $[a,b]$ one has
\begin{equation}
     \bar {\eusm O}[f_1,f_2,a,b] \equiv \frac{f_1 \cdot f_2}{|f_1|\,|f_2|} 
     \qquad\quad
     f_1 \cdot f_2 \equiv \int_{a}^{b} dx \, f_1(x) f_2(x)     
     \label{eq:40}
\end{equation}
but in case of fast-decaying functions (such as the correlators considered here) a measure 
that is more uniform over the domain and more stringent is provided by the 
``relative overlap'', namely
\begin{equation}
    {\eusm O}[f_1,f_2,a,b] \equiv \bar{\eusm O} \bigl[ 1,\frac{f_2}{f_1},a,b \bigr]
     \label{eq:60}     
\end{equation}
Note that for functions $f(x_k)$ defined only on discrete set of arguments $x_k$
(lattice correlators) we apply the above formulas as well via replacing these 
functions with their piecewise constant extensions $f(x)=f(x_k)$ for $x\in [x_k,x_{k+1})$.

\begin{table}[t]
  \centering
  \begin{tabular}{cccccc}
  \hline\\[-0.4cm]
  \multicolumn{1}{c}{ensemble}  &
  \multicolumn{1}{c}{$a$ [fm]}  &
  \multicolumn{1}{c}{$V$}  &
  \multicolumn{1}{c}{$V_p$ [fm$^4$]} &
  \multicolumn{1}{c}{$\quad$configs$\quad$} \\[2pt]
  \hline\\[-0.4cm]
   $\;\cE_1\;$ & $\quad 0.110 \quad$ & $\quad  12^4 \quad$ & $\quad 3.0 \quad$ 
           & $\quad 142 \quad$\\
   $\;\cE_2\;$ & $\quad 0.082 \quad$ & $\quad 16^4 \quad$ & $\quad 3.0 \quad$ 
           & $\quad 50 \quad$\\
   $\;\cE_3\;$ & $\quad 0.066 \quad$ & $\quad 20^4 \quad$ & $\quad 3.0 \quad$ 
           & $\quad 20 \quad$\\
   $\;\cE_4\;$ & $\quad 0.055 \quad$ & $\quad 24^4 \quad$ & $\quad 3.0 \quad$ 
           & $\quad 10 \quad$\\
\hline 
\end{tabular}
\caption{Ensembles of Iwasaki gauge configurations for topological density 
calculations.}
\label{ensemb_tab} 
\vskip -0.15in
\end{table}

\begin{figure}
\begin{center}
    \vskip -0.20in
    \centerline{
    \hskip -0.00in
    \includegraphics[width=11.0truecm,angle=0]{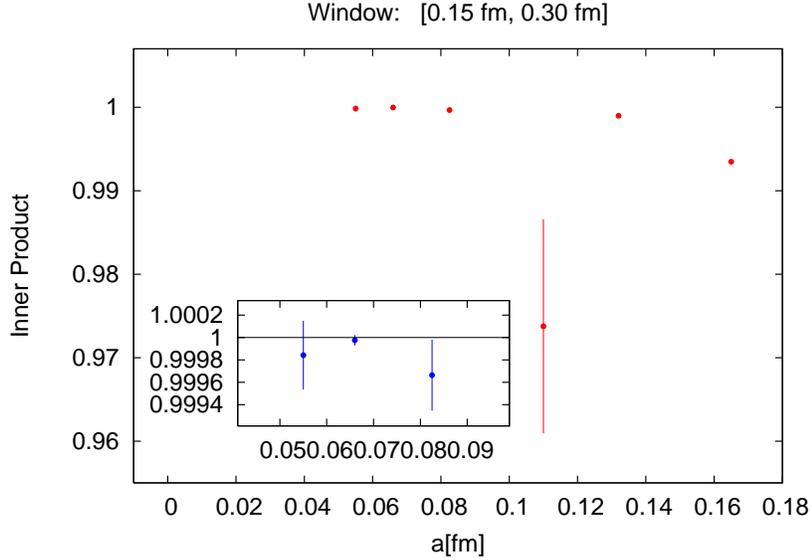}
     }
     \vskip -0.00in
     \caption{The relative overlap ${\eusm O}[\Gbar(a),G(a),0.15/a,0.30/a]$.
              See the discussion in the text for explanation.}
     \label{fig:adep}
     \vskip -0.42in 
\end{center}
\end{figure}

Consider the lattice definition of the topological density 2--point function in 
the domain of physical distances $[r_{p,1}, r_{p,2}]$. This 
requires studying the behavior of $G(r,a)$ on the ever--growing sliding lattice 
interval $[\,r_{p,1}/a,\,r_{p,2}/a\,]$ close to the continuum limit $a\to 0$.    
If the original and the geometric correlators ($G(r,a)$, $\Gbar(r,a)$) carry 
the same information about the shape of the physical 2-point function in 
the continuum limit, their relative overlap has to approach unity. The result 
of such calculation for the interval [0.15 fm, 0.30 fm] is 
shown in Fig.~\ref{fig:adep}. To understand this, we also included two ensembles 
coarser than $\cE_1$, with relevant lattice interval being entirely within 
the positive core of the correlator. For $\cE_1$ the upper limit of the interval 
involves a correlation very close to zero as the correlator just becomes negative. 
This results in the observed dip and large fluctuation of the relative overlap whose 
definition implicitly assumes that $f_1$ is non--zero on its domain 
(see Eq.~(\ref{eq:60})). For finer lattice spacings the relative overlap quickly
approaches unity. Given these results, we propose to consider 
the possibility that the following conjecture holds.
\smallskip

\noindent {\bf Conjecture 1.} {\em Consider the pure-glue lattice gauge theory defined by 
the Iwasaki gauge action on the infinite lattice together with arbitrary non--perturbative
procedure of fixing the lattice spacing $a$. If $G(a)$ and $\Gbar (a)$ are the original and 
the geometric 2--point functions of standard overlap--based topological density then 
\begin{equation}
     \lim_{a\to 0} {\eusm O} \bigl[\,\Gbar(a),\, G(a),\, \frac{r_{p,1}}{a},\, 
                                                         \frac{r_{p,2}}{a}\,\bigr]
     \,=\, 1
     \label{eq:80}  
\end{equation} 
for arbitrary range of physical distances $[r_{p,1}, r_{p,2}]$.}
\smallskip

\noindent We emphasize that we do not mean to imply that the simplest sign reduction
$q(x) \rightarrow \sgn(q(x))$ considered here is the unique way to obtain 
the above--proposed equivalence. Rather, what we wish to convey is that there exists 
a family of computable configuration--based reductions 
$\{\,q(x)\,\} \rightarrow  \{\,sq(x)\,\}$ with $sq(x)\in \{-1,1\}$ such that 
the shape equivalence holds.
\smallskip

\begin{figure}
\begin{center}
    \vskip -0.20in
    \centerline{
    \hskip -0.00in
    \includegraphics[width=11.0truecm,angle=0]{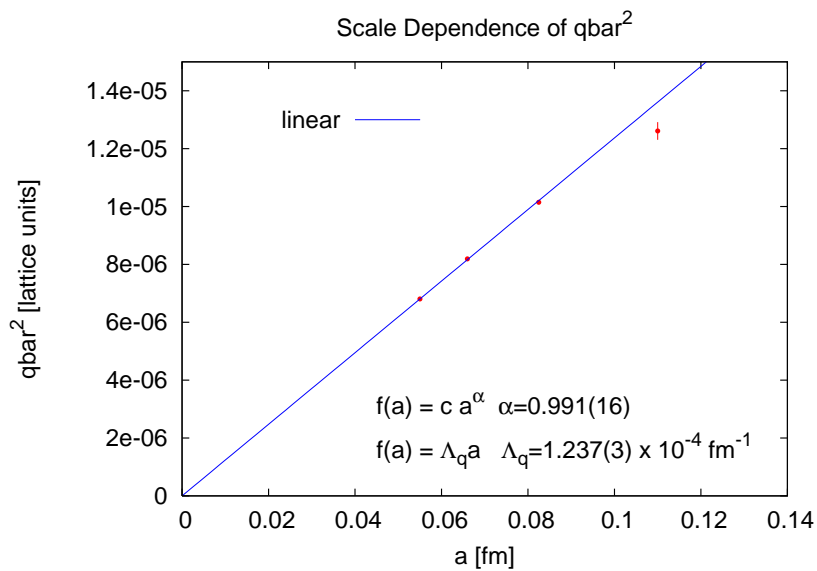}
     }
     \vskip -0.00in
     \caption{The scale dependence of $\qbar^2(a)$.See the discussion in the text for 
     explanation.}
     \vskip -0.40in 
     \label{fig:a_qbar}
\end{center}
\end{figure}

\noindent {\bf 4. The Uses and the Discussion.} Dominance of sign--coherent geometry
promises to have interesting ramifications regardless of whether it eventually turns out 
to be an approximation (exact at asymptotically large physical distances) or an exact 
statement specified by Conjecture 1. In the former case, it suggests to consider
the ``homogeneous approximation'' to fundamental topological structure~\cite{Hor08B}. 
In the latter case (strictly consistent with the data at this point) it would represent 
a simple explicit link between short and long--distance 
physics -- a connection that the fundamental structure (in any composite field) has to 
encode in some way. There are several possible ways to exploit such exact link, and we 
mention two examples here. (1) Validity of Conjecture 1 would imply that for spectral 
calculations (glueball masses) it is sufficient to evaluate configurations of signs of 
topological 
density rather than their precise values. For overlap--based topological density, this 
could lead to a speedup of calculations. (2) An alternative (not strictly equivalent)
way of formulating the observed trends is the following. {\em Consider 
$\lim_{r\to \infty} G(r,a)/\Gbar(r,a)$. This limit, $\qbar^2(a)$, exists and 
the correlators $G(r,a)$ and ${\hat G}(r,a)\equiv \Gbar(r,a) \qbar^2(a)$ are strictly 
equivalent from the point of view of the continuum limit.} In particular, in the expected 
way of obtaining finite physical 2-point function we have at arbitrary physical distance
$r_p$
\begin{equation}
    G_p(r_p) \equiv \lim_{a\to 0} \frac{G(\frac{r_p}{a},a)}{a^8} \,=\,
     \lim_{a\to 0} \frac{{\hat G}(\frac{r_p}{a},a)}{a^8} \,=\,
     \lim_{a\to 0} \frac{ \Gbar(\frac{r_p}{a},a) \, \qbar^2(a)}{a^8}
    \label{eq:100}  
\end{equation} 
Applying this equation to short distances $x_p \ll 1/m_p$ (glueball range) and taking 
into account 
that $\Gbar(r,a)$ has finite continuum limit at fixed $r$, one can see that measuring 
$\qbar^2(a)$ offers a novel way of testing the exact nature of asymptotic freedom. 
In Fig.~\ref{fig:a_qbar} we show the result of evaluating $\qbar^2(a)$ for ensembles 
$\cE_1$--$\cE_4$ (fitting $G(r,a)/\Gbar(r,a)$ to a constant), indicating that for our 
finest lattices (2.3--3.6 GeV) the behavior is excellently described by  
$\qbar^2(a)\propto a$. The perturbative prediction, namely a logarithmic decrease 
of $\qbar^2(a)$ toward zero, hasn't yet set in. It will be quite interesting to examine 
yet shorter scales to see where the perturbative description starts to dominate,
and to look for possible geometric changes occurring in the fundamental structure as that 
happens. The absence of such transition would imply a highly non--standard scenario, in which
$G_p(x_p)\propto 1/|x_p|^7$ at short distances with the true dimension of the topological 
density operator being $D_q=3.5$ rather than $D_q=4$. In that case the parameter $\theta$ 
would be dimensionful (analogous to mass), asymptotic freedom would be only approximate, 
and the strong CP problem would cease to exist.
\footnote{It is tempting to interpret $\qbar(a)$ as the density
of the underlying fundamental structure whose features are fully encoded in correlation 
functions at non--zero distances but distorted by short--range noise at zero physical 
distance. While exact nature of such interpretation depends on the details of the interplay 
between the noise and the structure, it might be sufficiently robust so that the behavior 
$\qbar(a) \propto a^{D_a}$ could serve as a definition of the ``analytic dimension''
($D_a$--dimensional local density). 
From Eq.~(\ref{eq:100}) one could then deduce the relation between the true dimension of 
the operator and the analytic dimension of the associated structure, namely $D_q+D_a=4$. 
Persistence of the trends shown in Fig.~\ref{fig:a_qbar} would then imply $D_a=1/2$, 
while the validity of perturbation theory at short distances means that $D_a=0$.}

\begin{thebibliography}{99}

\bibitem{Hor03A} I.~Horv\'ath \emph{et al}., Phys.~Rev. {\bf D68}, 114505 (2003).

\bibitem{Ale05A} A.~Alexandru, I.~Horv\'ath, J.B.~Zhang,
                 Phys. Rev. {\bf D72} (2005) 034506.

\bibitem{Hor06fram} I.~Horv\'ath, {\tt hep-lat/0605008}; 
                    {\tt hep-lat/0607031};
                    {\em ``Framework for Systematic Study of QCD 
                     Vacuum Structure III: Scale Dependence''}, in preparation.

\bibitem{Boy06A} P.Y. Boyko, F.V. Gubarev, 
                 Phys. Rev. {\bf D 73}, 114506 (2006).

\bibitem{Ilg07A} E.M.~Ilgenfritz, \emph{et al}., Phys. Rev. {\bf D76}, 034506 (2007). 

\bibitem{Hor05A} I.~Horv\'ath, \emph{et al}., Phys. Lett. {\bf B612} (2005) 21.

\bibitem{other_fun} 
         S.~Ahmad, J.T.~Lenaghan, H.B.~Thacker, Phys. Rev. {\bf D72}, 114511 (2005);
         H.B.~Thacker, PoS {\bf LAT2005} (2005) 324, {\tt hep-lat/0509057};
         F.L.~Lin, S.Y.~Wu, {\tt arXiv:0805.2933};
         P.J.~Moran, D.B.~Leinweber, {\tt arXiv:0805.4246}.

\bibitem{Hor08B} I.~Horv\'ath, A.~Alexandru, T.~Streuer, in preparation.

\bibitem{Neu98BA}
   H.~Neuberger, Phys.~Lett. {\bf B417} (1998) 141; 
                 Phys.~Lett. {\bf B427} (1998) 353.

\bibitem{Chen03} Y.~Chen \emph{et al}., Phys.~Rev. {\bf D70}, 034502 (2004).


\end{thebibliography}
\end{document}